\documentclass[12pt]{article}
\usepackage{siunitx}
\usepackage{graphicx}
\usepackage[utf8]{inputenc}
\usepackage{graphicx,xcolor,setspace,geometry}
\usepackage{amsmath,setspace}
\usepackage{amssymb}
\usepackage{mathtools}
\usepackage{multirow}
\usepackage{booktabs}
\usepackage{threeparttable}
\usepackage{adjustbox}
\usepackage{nicefrac}
\usepackage{natbib}
\usepackage{multirow}
\usepackage{color}
\usepackage{caption}
\usepackage{hyperref}

\title{Maximum approximate likelihood estimation of general continuous-time state-space models}

\author{Sina Mews$^{1,}\footnote{Corresponding author; email: \texttt{sina.mews@uni-bielefeld.de}.}$, Roland Langrock$^{1}$, Marius Ötting$^{1}$, \\ Houda Yaqine$^{1}$, and Jost Reinecke$^{2}$ \\
 \\
$^{1}$Department of Business Administration and Economics, \\ Bielefeld University, Germany \\
$^{2}$Faculty of Sociology, Bielefeld University, Germany}

\date{}

\begin{document}

\begin{spacing}{1.25}
    \maketitle
\end{spacing}

\vspace{-5mm}

\begin{spacing}{1.5}

\begin{abstract}
Continuous-time state-space models (SSMs) are flexible tools for analysing irregularly sampled sequential observations that are driven by an underlying state process. Corresponding applications typically involve restrictive assumptions concerning linearity and Gaussianity to facilitate inference on the model parameters via the Kalman filter. 
In this contribution, we provide a general continuous-time SSM framework, allowing both the observation and the state process to be non-linear and non-Gaussian.
Statistical inference is carried out by maximum approximate likelihood estimation, where multiple numerical integration within the likelihood evaluation is performed via a fine discretisation of the state process.
The corresponding reframing of the SSM as a continuous-time hidden Markov model, with structured state transitions, enables us to apply the associated efficient algorithms for parameter estimation and state decoding. 
We illustrate the modelling approach in a case study using data from a longitudinal study on delinquent behaviour of adolescents in Germany, revealing temporal persistence in the deviation of an individual's delinquency level from the population mean.
\end{abstract}


\section{Introduction}

State-space models (SSMs) are flexible tools for analysing sequential observations that depend on underlying non-observable states, with interest and hence inference typically centred on the states.
There are two main conceptual decisions to be made when tailoring an SSM to any given application, concerning a) the nature of the state space and b) whether the state process is defined as operating in discrete or continuous time.
Regarding a), the nature of the state space depends on the interpretation of the latent variable. 
The latter could relate either to  discrete states, for example indicating an individual's health status \citep[e.g.\ infected vs.\ not infected;][]{conn2009} or an animal's behavioural modes \citep[e.g.\ travelling, resting, and foraging;][]{vanBeest2019}, or to continuous states, for example related to the nervousness of the financial market \citep[e.g.\ within stochastic volatility models;][]{kim1998stochastic} or to an athlete's current form \citep[e.g.\ in analyses of serial correlation in performance;][]{oetting2020}. 
In some applications, the specification of the state space is obvious (e.g.\ in simple capture-recapture studies, with states corresponding to dead and alive; \citealp{king2016semi}), whereas in others it constitutes a modelling choice (e.g.\ in stochastic volatility modelling, where the market states are commonly considered to be continuous, but sometimes dichotomised to calm and nervous, respectively; \citealp{bulla2006stylized}).
Regarding b), i.e.\ the decision whether the model is defined as operating in discrete or continuous time, the time formulation is usually determined by the sampling scheme of the data at hand. 
While discrete-time models are appropriate for time series with regular time intervals, continuous-time models are more suitable for irregularly spaced observations.
However, as irregularly sampled data can often be augmented via imputation to give a regular series, or temporarily aggregated to yield regularly spaced observations, the choice of the time formulation is not necessarily trivial.

With these two dimensions along which a conceptual modelling decision needs to be made, we distinguish four possible formulations of state-space models as presented in Table~\ref{tab:decisions}, with either discrete or continuous states and operating in either discrete or continuous time.
We refer to models with finite state space as hidden Markov models (HMMs), using the label SSM to refer to models with infinitely many and usually continuous-valued states \citep[a distinction commonly made in the literature, though some authors refer to both model classes as HMMs; e.g.][]{cappe2005}.
In terms of statistical inference, discrete-time HMMs arguably constitute the simplest case from Table~\ref{tab:decisions}. 
In particular, for these models there are recursive schemes, e.g.\ to evaluate the likelihood, that are applicable under various dependence structures and flexible distributional assumptions.
The continuous-time formulation of HMMs is only slightly more involved, with the underlying state process then governed by a continuous-time (rather than a discrete-time) Markov chain.
The main inferential tools available for discrete-time HMMs are applicable also in continuous time \citep{jackson2003}, though some extensions, e.g.\ to accommodate time-varying covariates, are not straightforward anymore \citep[e.g.][]{michelot2019, mews2020, williams2020}. 
When considering a model in discrete time but with a continuous state space, i.e.\ an SSM, then inference is straightforward only in the linear and Gaussian case, for which the Kalman filter is applicable \citep[e.g.][]{mccrea2010, durbin2012}. 
In the more general case, inference is hindered by the fact that the likelihood contains multiple integrals, making direct evaluation difficult \citep[e.g.][]{kitagawa1987, bartolucci2003, langrock2011}.

\begin{table}[!htb]
    \centering
    \begin{tabular}{ll|cc}
     & & \multicolumn{2}{c}{latent variable (state)} \\
     & & discrete & continuous \\ 
     \hline
     & discrete & HMMs & SSMs \\[-0.6em]
     time & & & \\[-0.6em]
     & continuous & continuous-time HMMs & continuous-time SSMs
\end{tabular}
    \caption{Possible formulations of state-space models.}
    \label{tab:decisions}
\end{table}

Despite these difficulties that arise when extending (discrete-time) HMMs \textit{either} to have a continuous state space \textit{or} to be formulated in continuous time, the corresponding extensions are nevertheless well covered in the existing literature and are fairly routinely applied. 
In this paper, we focus on the fourth case from Table \ref{tab:decisions}, i.e.\ SSMs that are formulated in continuous time (and are not necessarily linear and Gaussian). 
Such models, which are not nearly as well documented in the literature as the other three classes from Table \ref{tab:decisions}, are relevant in the context of irregularly sampled data in conjunction with an underlying continuous-valued state process.
In particular, irregularly spaced observations are quite common in data sets on natural phenomena such as earthquakes \citep[e.g.][]{beyreuther2008}, in medical data \citep[e.g.][]{amoros2019}, or in survey data, which for example relate to psychological measurements \citep[e.g.][]{oravecz2011}.
While continuous-time modelling can sometimes be avoided also in case of irregular sampling, for example using imputation methods as in \citet{kim2008}, continuous-time SSMs are more realistic and flexible than models that assume simplifications of either the time formulation or the nature of the latent variable. 
In some applications, continuous-time SSMs with a diffusion state process are considered (see, e.g., \citealp{niu2016,lavielle2018,michelot2020}), but except for \citet{albertsen2015}, who use t-distributed measurement errors, both the state and the observation process are usually assumed to be linear and Gaussian to allow for the application of the Kalman filter (see, e.g., \citealp{johnson2008,tandeo2011,dennis2014,koopman2018, jonsen2020}).

In our contribution, we present a flexible framework for continuous-time SSMs, allowing both the observation process as well as the state process to be non-linear and non-Gaussian.
Our approach thus enables a variety of possible model specifications, requiring only that the transition density of the state process has an explicit analytic form.
The latter condition is satisfied by all linear processes, including the Ornstein-Uhlenbeck (OU) process, as well as the non-linear geometric Brownian motion and the Cox-Ingersoll-Ross process.
As the model's likelihood involves intractable integration over all possible realisations of the continuous-valued state process at each observation time, we follow ideas from \citet{kitagawa1987}, \citet{bartolucci2003}, and \citet{langrock2011} and approximate the integral by finely discretising the state space.
This approximation can be regarded as a reframing of the model as a continuous-time HMM with a large but finite number of states, enabling us to apply the corresponding efficient algorithms.
Moreover, transferring our model to an HMM framework not only allows to specify any non-linear function in the observation process, but also to select any distribution for the error terms.
Therefore, the main advantage of our approach is its great flexibility to easily consider non-linear and non-Gaussian observations with underlying continuous-valued variables, whereas most continuous-time SSMs in the literature are bound to specific applications.

In Section \ref{sec:methods}, we first discuss statistical inference for continuous-time SSMs based on approximating the likelihood via state discretisation.
Subsequently, in Section \ref{sec:sim}, we demonstrate the feasibility of our approach and investigate the estimation accuracy in simulation experiments. 
An illustrating case study on delinquent behaviour of adolescents is presented in Section \ref{sec:applic}.

\section{Methodology}
\label{sec:methods}

We consider a sequence of random variables, $Y_{t_1},\ldots,Y_{t_T}$, observed at discrete but irregularly spaced time points $t_0,t_1,\ldots,t_T$, where $0=t_0<t_1<\ldots<t_T$. 
This observation process is assumed to be driven by an underlying and non-observable continuous-valued state process, $\{X_t\}_{t \geq 0}$, which operates in continuous time and is assumed to be Markovian. 
The distribution of $Y_{t_k}$, $k=1,\ldots,T$, is assumed to be fully determined by the underlying state $S_{t_k}$. 
In particular, the observations are assumed to be conditionally independent of each other, given the states. 
The model is thus specified via the conditional distributions
\begin{equation}
\label{SSMequation}
Y_{t_k} \,|\, X_{t_k} \quad \text{and} \quad X_{t_k} \,|\, X_{t_{k-1}}. 
\end{equation}
For the state process it is further assumed that the transition density, i.e.\ the probability density function of $X_{t}$ given $X_s=x_s$, for $t>s$, is available in closed form. 
In the following, this transition density is denoted by $p_{\Delta}(x_{t} | x_s)$ with $\Delta = t-s$. 
No restrictive assumptions are made for $Y_{t_k} \,|\, X_{t_k}$, specifically allowing the conditional distribution of the observations to be either continuous or discrete (and even categorical).

One possible choice for the state process is the OU process, which is described by the stochastic differential equation (SDE)
\begin{equation}
    d X_t = \theta (\mu - X_t) dt + \sigma d W_t, \quad X_0=x_0,
\label{OUprocess}
\end{equation}
where $\theta > 0$ is the drift parameter indicating the strength of reversion to the long-term mean $\mu \in \mathbb{R}$, $\sigma > 0$ controls the strength of fluctuations, and $W_t$ denotes the Brownian motion.
Due to its mean-reverting property, the OU process is a natural candidate for applications in which the latent variable fluctuates around some equilibrium state.

For simplicity of notation, we let $\tau = 0,1,\ldots,T$ denote the \textit{number} of the observation in the time series, such that in the following we use the shorthand notation $Y_\tau$ to indicate $Y_{t_\tau}$, and likewise $X_\tau$ for $X_{t_\tau}$, whenever unambiguous.
While $\tau$ is an integer, $t_\tau$ can be any non-negative number and represents the \textit{time} at which the observation $\tau$ was collected. 
Consequently, $\Delta_\tau  = t_\tau - t_{\tau-1}$ denotes the time difference between consecutive observations.

The likelihood of an SSM as in (\ref{SSMequation}) can be calculated by integrating over all possible values of the state process potentially underlying each observation time, resulting in an expression involving $T+1$ integrals.
To evaluate this multiple integral and hence the likelihood, we finely discretise the continuous-valued state space, as first proposed by \citet{kitagawa1987}.
Specifically, we define a range of possible values of the state process, $[b_0, b_m]$, which we divide into $m$ intervals $B_i=(b_{i-1}, b_i)$, $i=1,\ldots,m$, of equal length $(b_m-b_0) / m$, requiring both the range $[b_0, b_m]$ and $m$ to be sufficiently large.
Making use of the model's dependence structure and applying numerical integration, the SSM likelihood can then be approximated in the following way:
\begin{equation}
\begin{split}
    \mathcal{L}_T &= \int \ldots \int p(y_0, \ldots, y_T, x_0, \ldots, x_T) dx_T \ldots dx_0 \\
    &= \int \ldots \int p(x_0) p(y_0|x_0) \prod_{\tau=1}^T p_{\Delta_\tau} (x_\tau | x_{\tau-1}) p(y_\tau|x_\tau) dx_T \ldots dx_0 \\
    &\approx \sum_{i_0=1}^m \ldots \sum_{i_T=1}^m p(x_0 \in B_{i_0}) p(y_0| x_0=b^*_{i_0}) \\
    & \quad \times \prod_{\tau=1}^T p_{\Delta_\tau} (x_\tau \in B_{i_\tau} | x_{\tau-1} = b^*_{i_{\tau-1}}) p(y_\tau|x_\tau = b^*_{i_\tau}),
\label{likelihoodApprox}
\end{split}
\end{equation}
with $b_i^*$ denoting the midpoint of the interval $B_i$ and using $p$ as a general symbol for either a density or a probability.
There are alternative ways to approximate the multiple integral \citep[see, e.g.,][]{bartolucci2003, zucchini2016}, but which of these is used does not make a difference in practice, provided $m$ is sufficiently large.

The discretisation of the state space into $m$ intervals effectively amounts to an approximation of the SSM by an $m$-state HMM, which allows us to apply the entire HMM methodology to our model.
In particular, we can use the HMM forward algorithm to more efficiently calculate the approximate likelihood in Equation (\ref{likelihoodApprox}), which as it stands has a  computational cost of order $\mathcal{O}(Tm^T)$.
To recognise the approximation as an HMM, we specify the initial distribution $\boldsymbol{\delta} = (\delta_1, \ldots, \delta_m)$ with $\delta_i = p(x_0 \in B_i)$, and define the $i$-th entry of a diagonal matrix $\mathbf{P}(y_\tau)$ as $p(y_\tau | x_\tau = b^*_i)$.
Further, we define the $m \times m$ transition probability matrix $\Gamma_{\Delta_\tau} = (\gamma_{i,j}^{\Delta_\tau})$ by specifying $\gamma_{i,j}^{\Delta_\tau} = p(x_\tau \in B_j | x_{\tau-1} = b^*_i)$.
As indicated by the corresponding superscript, the state transition probabilities $\gamma_{i,j}^{\Delta_\tau}$ depend on the time difference $\Delta_\tau$ between consecutive observations. 
Using the HMM forward algorithm to calculate the approximate likelihood in Equation (\ref{likelihoodApprox}) reduces the computational cost to order $\mathcal{O}(Tm^2)$ and yields the following matrix product:
\begin{equation}
    \mathcal{L}_T \approx \boldsymbol{\delta} \mathbf{P}(y_0) \Bigl( \prod_{\tau=1}^T \boldsymbol{\Gamma}_{\Delta_\tau} \mathbf{P}(y_\tau) \Bigr) \boldsymbol{1},
\label{llk}
\end{equation}
where $\boldsymbol{1} \in \mathbb{R}^m$ denotes a column vector of ones.

The entries in $\mathbf{P}(y_\tau)$ are simply the conditional densities or probabilities as determined by the model assumed for the observation process, $Y_{\tau} \,|\, X_{\tau}$ (a concrete example will be given in Section \ref{sec:applic}). To illustrate how $\Gamma_{\Delta_\tau}$ is obtained, consider the example of the OU process in Equation (\ref{OUprocess}). This process has a Gaussian transition density $p_\Delta(x_\tau | x_{\tau-1}) $,
such that 
\begin{equation*}
    X_\tau | X_{\tau-1} = x \sim \mathcal{N}\left( \text{e}^{-\theta \Delta_\tau} x + \mu \bigl(1- \text{e}^{-\theta \Delta_\tau}\bigr), \quad 
    \frac{\sigma^2}{2\theta} \bigl(1- \text{e}^{-2\theta \Delta_\tau}\bigr) \right), 
\end{equation*}
based on which the transition probabilities $\gamma_{i,j}^{\Delta_\tau}$ can be calculated \citep[see, e.g.,][]{cerbone1981}.
For this state process and assuming stationarity, the initial state probabilities $\delta_i$ can be calculated based on the limiting distribution $ X_\tau \sim \mathcal{N} \left(\mu, \frac{\sigma^2}{2\theta} \right) $ of the OU process.

Irrespective of the specific assumptions made for $Y_{\tau} \,|\, X_{\tau}$ and $ X_{\tau} \,|\, X_{\tau-1}$, the parameters of the SSM can be estimated by numerically maximising the approximate likelihood in Equation (\ref{llk}), subject to standard technical issues as detailed for example in \citet{zucchini2016}. In practice, the range of the state process $[b_0, b_m]$ as well as the number of intervals $m$ used for the likelihood approximation need to be specified.
Regarding the choice of $[b_0, b_m]$, it is important to cover the essential range of possible values of the state process, which can be examined by looking at the (estimated) stationary distribution of the state process, if available. 
For example, for the OU process, a conservative choice would be $\bigl[-3\sigma^2/\theta,3\sigma^2/\theta\bigr]$ (corresponding to six times the standard deviation in either direction).
Regarding the choice of $m$, it is intuitively clear that the more intervals are used, the closer the likelihood can be approximated, but the longer the computation time --- a classical trade-off situation.
Therefore, to gain some understanding of how many intervals are sufficient for the likelihood approximation, the next section will investigate the effect of $m$ on the estimation accuracy.

\section{Simulation experiments}
\label{sec:sim}

Simulations were conducted to explore the effect of approximating the likelihood by discretising the continuous-valued state process, in particular with regard to the estimation accuracy. 
While the likelihood approximation can be rendered arbitrarily accurate by using increasingly many intervals in the discretisation, it is not clear at which number of intervals $m$ the parameter estimation stabilises such that increasing $m$ does not (substantially) change the estimation results anymore.
We further investigate if the appropriate number of intervals $m$ needed for the approximation depends on the variability of the underlying state process.

\begin{figure}[!tbh]
    \centering
    \includegraphics[width=150mm]{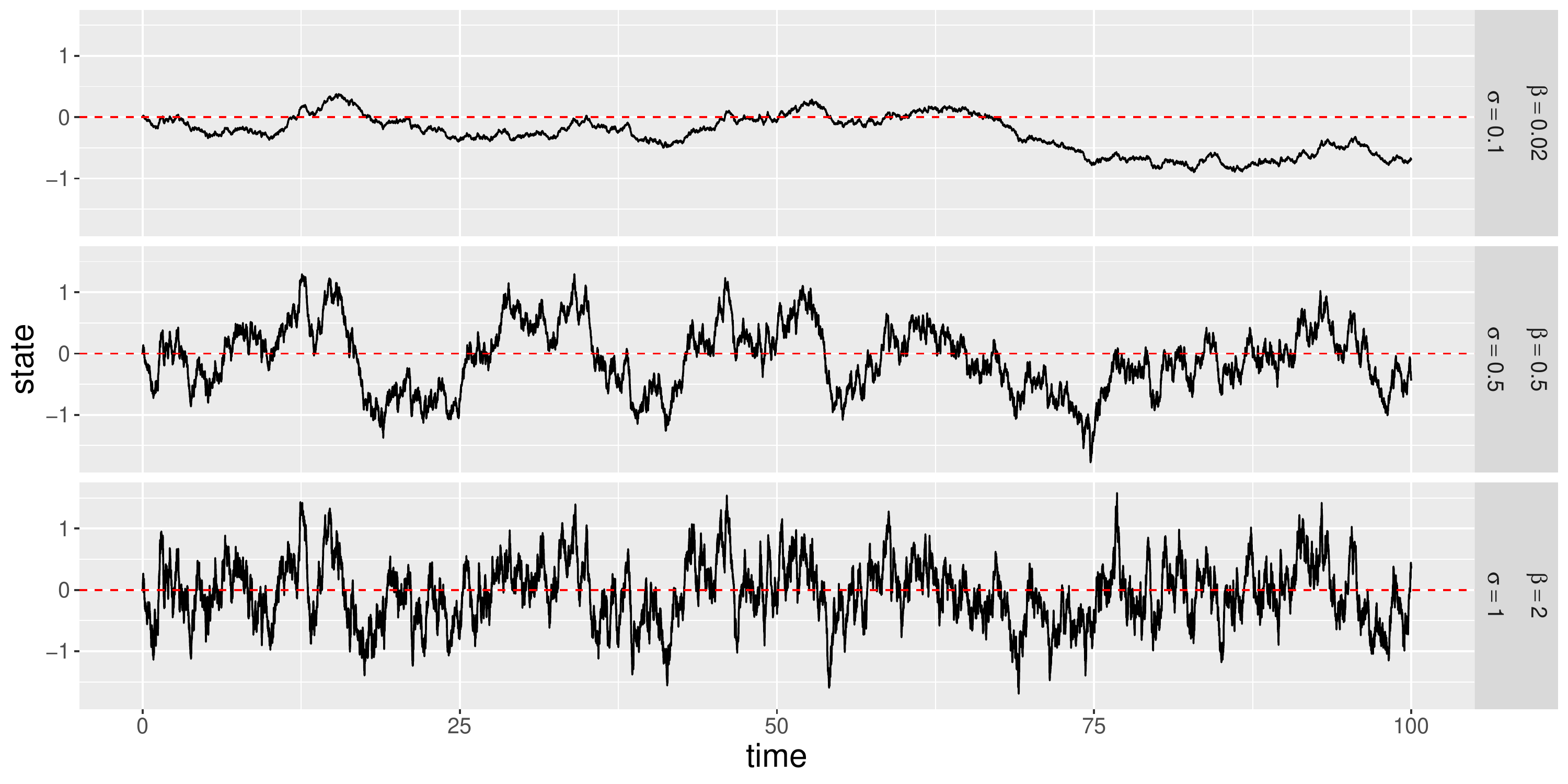}
    \caption{Example path realisations of the OU processes considered. The graphs were obtained by application of the Euler-Maruyama scheme with initial value 0 and step length 0.01.}
    \label{fig:OUprocesses}
\end{figure}

We consider three simulation settings, in which the state process is modelled using the OU process (cf.\ Equation (\ref{OUprocess})) with long-term mean $\mu = 0$. 
For the drift term $\theta$ and the diffusion parameter $\sigma$, we choose four different parameter combinations, $(\theta,\sigma) \in \{ (0.02,0.1), (0.5,0.5), (2,1) \}$, which all share the same limiting distribution, namely $ X_\tau \sim \mathcal{N} \left(0, 0.5^2 \right) $.
The variability, as governed by the diffusion parameter $\sigma$, increases from Setting 1 to Setting 3.
Example path realisations of the three state processes considered are shown in Figure~\ref{fig:OUprocesses}.
The observation process is assumed to be a Poisson-distributed, irregularly spaced sequence of counts with
\begin{equation*}
    Y_\tau \sim \text{Poisson}(\lambda_\tau), \quad
        \lambda_\tau = \text{exp}(X_\tau) \alpha,
\end{equation*}
such that the mean of the observation process fluctuates (asymmetrically) around $\alpha>0$.
We set $\alpha = 200$ and generate one sequence of $T = 2000$ observations for each setting. 
The time intervals between consecutive observations are measured in days and were drawn from a Poisson distribution with a mean of 30 hours (the time scales are arbitrary here and are stated merely to aid interpretation).
The simulated count data for each setting are shown in Figure~\ref{fig:simData} in the Appendix.

For parameter estimation, we approximate the likelihood by discretising the state space as described in Section \ref{sec:methods}, and vary the number of intervals $m$ used in the approximation.
In each setting, we thus repeatedly estimate the model parameters for a \textit{single} sequence of counts by numerically maximising the likelihood given in Equation (\ref{llk}), considering $m = 20,30,50,100,150$ intervals and choosing a range of $[b_0=-2.5, b_m=2.5]$ for the state process.

\begin{table}[]
    \centering
    \begin{tabular}{lccccl} \hline
         \multicolumn{6}{c}{Setting 1}  \\ \hline
         & $\theta$ & $\sigma$ & $\alpha$ & comp.\ time (sec) & -- llk \\
         $m=20$ & 0.0167 & 0.106 & 285.3 & 6.1 & 9874.21 \\
         $m=30$ & 0.0186 & 0.096 & 177.9 & 8.9 & 9577.25 \\
         $m=50$ & 0.0164 & 0.098 & 167.5 & 18.1 & 9542.46 \\
         $m=100$ & 0.0174 & 0.101 & 191.4 & 37.1 & 9544.06 \\
         $m=150$ & 0.0174 & 0.101 & 191.5 & 65.7 & 9544.15 \\
         true values & 0.02 & 0.1 & 200 & & \\ \hline
         \multicolumn{6}{c}{Setting 2}  \\ \hline
         & $\theta$ & $\sigma$ & $\alpha$ & comp.\ time (sec) & -- llk \\
         $m=20$ & 0.190 & 0.453 & 339.6 & 4.8 & 12009.62 \\
         $m=30$ & 0.484 & 0.495 & 211.5 & 7.4 & 11722.41 \\
         $m=50$ & 0.494 & 0.499 & 193.7 & 17.2 & 11715.25 \\
         $m=100$ & 0.495 & 0.500 & 197.7 & 26.3 & 11715.03 \\
         $m=150$ & 0.495 & 0.500 & 197.7 & 41.5 & 11715.04 \\
         true values & 0.5 & 0.5 & 200 & & \\ \hline
         \multicolumn{6}{c}{Setting 3}  \\ \hline
         & $\theta$ & $\sigma$ & $\alpha$ & comp.\ time (sec) & -- llk \\
         $m=20$ & 2.345 & 1.080 & 195.5 & 8.4 & 12275.82 \\
         $m=30$ & 2.343 & 1.086 & 194.0 & 7.9 & 12083.40 \\
         $m=50$ & 2.361 & 1.090 & 198.4 & 13.8 & 12065.85 \\
         $m=100$ & 2.358 & 1.091 & 199.1 & 35.2 & 12066.90 \\
         $m=150$ & 2.358 & 1.091 & 199.1 & 64.6 & 12066.90 \\
         true values & 2 & 1 & 200 & & \\ \hline
    \end{tabular}
    \caption{Estimated parameters, computation times and maximum log-likelihood (llk) values in the simulation experiments considering different numbers of intervals $m$ used in the likelihood approximation.}
    \label{tab:simRes}
\end{table}

For each simulation setting and the different numbers of intervals $m$ considered, the maximum log-likelihood values, the relative biases of the estimated parameters, and the computation times are shown in Table~\ref{tab:simRes}.
The computation time increases with increasing interval numbers, whereas the maximum likelihood values as well as the estimated parameters stabilise with increasing $m$.
This is to be expected: given a sufficiently fine discretisation, a further increase in the interval numbers does not yield a relevant difference in the likelihood value.
There are, however, some differences between the three settings considered: 
while in Setting 1, the estimation results stabilise not until $m \geq 100$, the likelihood values and the estimated parameters do not change much in the other settings when increasing the number of intervals to $m>50$.
The more variable the underlying OU process (i.e.\ the larger the diffusion $\sigma$), the less intervals $m$ are thus needed in the approximation.
In other words, when the observations fluctuate considerably, then the discretisation of the state process does not need to be as fine as when the process has a higher persistence, in which case a finer discretisation is required to detect the associated more gradual changes.
A conservative choice of $m \geq 100$ is however advisable, provided that the resulting computational cost is acceptable.
In our simulations, fitting the models with $m \geq 100$ took about one minute on a 1.6 GHz Intel\textsuperscript{\textregistered} Core\textsuperscript{TM} i5 CPU.

In a second simulation experiment, we ran an empirical check of the estimators' consistency.
Specifically, focusing on Setting 2 above, we simulated 200 data sets of $T = 2000, 5000, 10000$ observations each and then estimated the model parameters with fixed $m = 100$.
The results indicate that in this particular setting, the parameter estimators are approximately unbiased already for $T = 2000$, with the precision increasing with increased sample size (see Figure~\ref{fig:bias} in the Appendix).

\section{Case study on delinquent behaviour in adolescence and young adulthood} 
\label{sec:applic}

\subsection{Model formulation}

We analyse data from the longitudinal research project \textit{Crime in the Modern City} on deviant and delinquent behaviour of adolescents and young adults in Western Germany \citep[for more details see][]{boersetal2010, seddig2017}. 
The survey was first conducted in the year 2000 and comprised students in the 7th grade at public schools, who were mostly 12 to 13 years old. 
This cohort was repeatedly interviewed by means of self-administered questionnaires over a study period of 16 years.
In each survey, the participants were asked about various offences like graffiti spraying, shop-lifting, drug abuse, or assault with and without a weapon, and indicated how often they had committed each offence in the twelve months prior to the survey. 
The data collection, however, did not follow a regular sampling scheme as the first eight waves of the panel study were administered annually, while the last four waves were conducted biannually.
Further, due to wave nonresponse, meaning that some participants would not respond in one or more panel waves, the data set contains missing values, which is quite common in longitudinal studies.
As a consequence, the length of time intervals between consecutive observations is irregular and ranges from one to four years. 

\begin{figure}[!htb]
    \centering
    \includegraphics[width=0.95\textwidth]{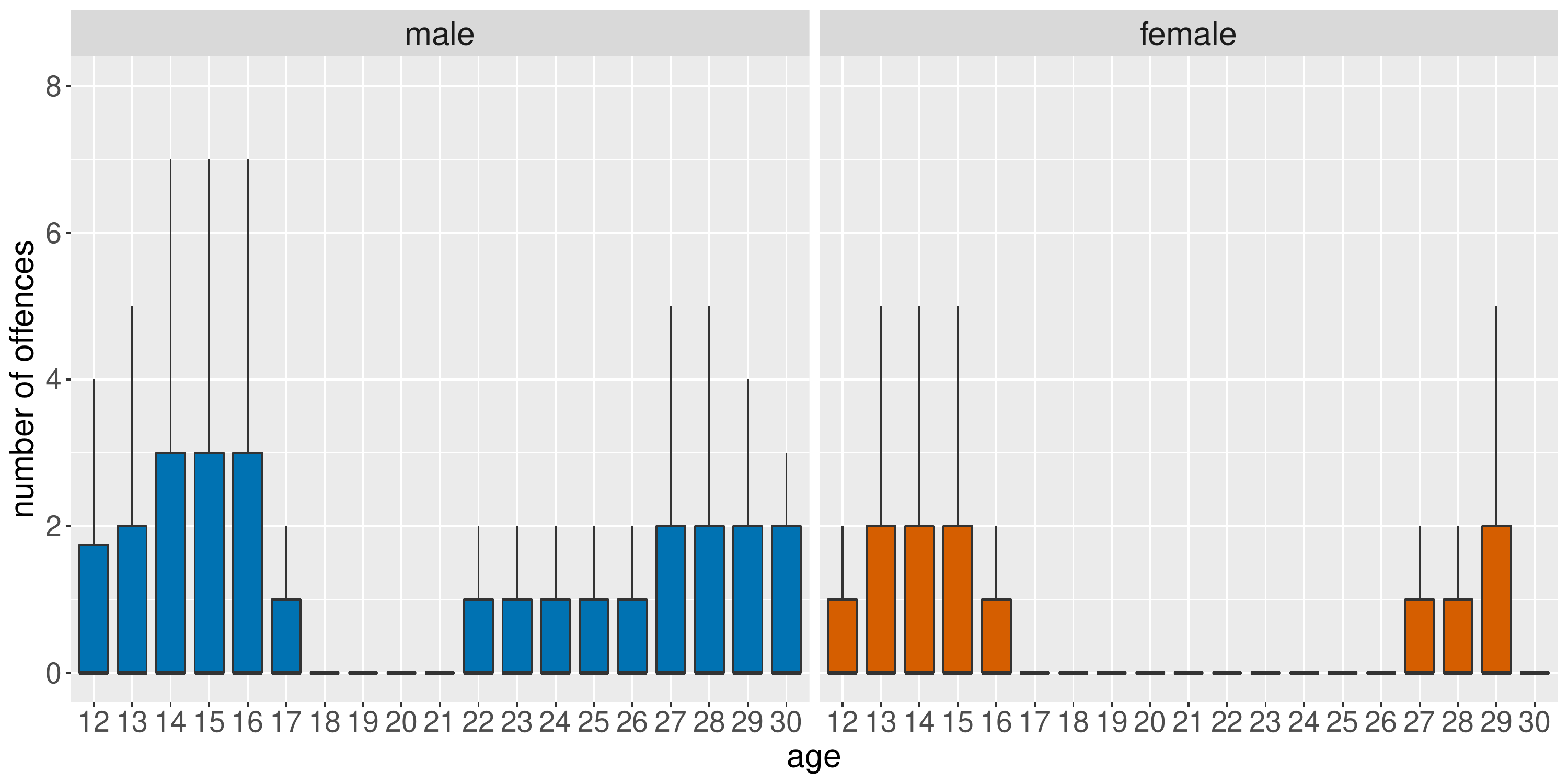}
    \caption{Boxplots of the number of offences committed in the twelve months prior to the survey for different age classes and both gender. Outliers have been removed from the plot for clarity. The same figure including outliers is provided in Figure~\ref{fig:boxplots_outliers} in the Appendix.}
    \label{fig:boxplots}
\end{figure}

In this case study, we consider the total number of offences indicated in each survey, from which individual trajectories of delinquent behaviour can be constructed. 
We included all participants who committed at least one offence within the study period, resulting in 12327 observations from 1093 adolescents (467 male and 626 female).
The distribution of the number of offences for different age classes and both gender is shown in Figure~\ref{fig:boxplots}.
No delinquent behaviour was most often reported (72.6\% of observations), while overall the median number of offences, given that any were committed within the previous twelve months, is 3 (min: 1; max: 160).

The main aim is to investigate the persistence of the delinquency level, which is assumed to be a latent trait underlying the observed trajectories of adolescents' and young adults' delinquent behaviour. 
Therefore, we model the number of offences using an SSM, which we formulate in continuous time to address the irregular spacing of the observations as caused by the study design and the missing data.
Arguably, the data could also be regarded as a yearly time series with missing data and hence modelled using a discrete-time process --- however, a continuous-time process constitutes a convenient alternative, which directly accommodates the time gaps.
To allow for possible overdispersion, we assume the number of offences to follow a negative binomial distribution (conditional on the states).
As the study participants' age and gender are known to affect their delinquent behaviour \citep[e.g.][]{reinecke2013}, we additionally include these covariates in the observation process.
The observation process of the SSM is then specified as
\begin{equation}
\begin{split}
    Y_\tau \sim \hspace{1mm} &\text{NegBinom}(\nu_\tau, \phi), \\
        &\nu_\tau = \text{exp} \bigl(X_\tau + f_1(\text{age}_\tau) + f_2(\text{age}_\tau) \cdot \text{gender}_\tau \bigr), 
    \label{eq:crimeData}
\end{split}
\end{equation}
where $\nu_\tau$ is the mean of observation number $\tau$ (at time $t_\tau$), and $\phi$ is the dispersion parameter of the negative binomial distribution.
The mean is modelled as a function of the current state $X_\tau$ (i.e.\ the current delinquency level relative to the population mean for the relevant age group) as well as the covariate age and its interaction with gender. 
To allow for a nonlinear relationship, as indicated by Figure~\ref{fig:boxplots}, the effect of age on the number of offences committed is modelled nonparametrically.
Specifically, we set $f_i(\text{age}_\tau) = \sum_{l=1}^8 \omega_{i,l} C_l(\text{age}_\tau)$, for $i=1,2$, using cubic B-spline basis functions $C_l$ and 12 equally spaced knots ranging from 7 to 35 \citep[][]{deBoor1978, eilers1996}.
We further specify the state process to be an OU process with $\mu = 0$ (cf.\ Equation (\ref{OUprocess})), such that an individual's delinquency level --- or, more precisely, the deviation of the individual's delinquency level from the population mean --- is persistent over time and changes gradually.
Negative values of the state process then indicate that the individual is less inclined to delinquent activities, given its gender and age, whereas positive values indicate a higher inclination than would be expected based on gender and age.
The parameters of interest, i.e.\ the drift parameter and diffusion coefficient of the OU process for the state process, as well as the regression coefficients of the covariate effects and the dispersion parameter of the negative binomial distribution for the observation process, are estimated using maximum (approximate) likelihood as
described in Section \ref{sec:methods}. 
For the state discretisation, we set $m = 100$ and choose $[b_0=-9, b_m=9]$ as a possible range for the state process.
To assess whether the SSM formulation is actually needed to describe the structure in the data, we additionally fit a model without an underlying state process to the observations. 
This benchmark model is formulated according to Equation (\ref{eq:crimeData}), omitting $X_{\tau}$, and corresponds to the assumption that an individual's delinquency level is not persistent over time.

\subsection{Results}

According to the AIC, the continuous-time SSM is clearly favoured over the benchmark model without any state process ($\Delta$AIC: 1372).
The parameter estimates associated with the OU process and the dispersion parameter of the negative binomial distribution are shown in Table \ref{tab:estParams}.
Regarding the dispersion parameter, its small value reflects the large variation in the number of offences.
For the state process, the small value of the estimated drift parameter indicates fairly strong serial dependence, while the estimated diffusion coefficient shows that the deviations from zero can be large. 
In particular, the limiting distribution of the OU process is estimated as $ X_\tau \sim \mathcal{N} \left(0, 2.23^2 \right) $, indicating that considerable differences in the delinquency levels of adolescents can be observed over time.
This difference in and temporal persistence of latent delinquency levels can also be illustrated using simulated state trajectories based on the estimated parameters of the OU process (cf.\ Figure~\ref{fig:OUresults} in the Appendix).

\begin{table}[!htb]
    \centering
    \begin{tabular}{ccc}
        parameter & estimate & 95\% CI \\ \hline
        $\theta$ & 0.222 & [0.194; 0.255] \\
        $\sigma$ & 1.489 & [1.346; 1.647] \\
        $\phi$ & 0.570 & [0.483; 0.674]
    \end{tabular}
    \caption{Parameter estimates with 95\% confidence intervals (CIs) for the drift parameter $\theta$ and the diffusion coefficient $\sigma$ of the OU process as well as the dispersion parameter $\phi$ of the negative binomial distribution. The CIs were calculated based on the observed Fisher information.}
    \label{tab:estParams}
\end{table}

\begin{figure}[!htb]
    \centering
    \includegraphics[scale=0.4]{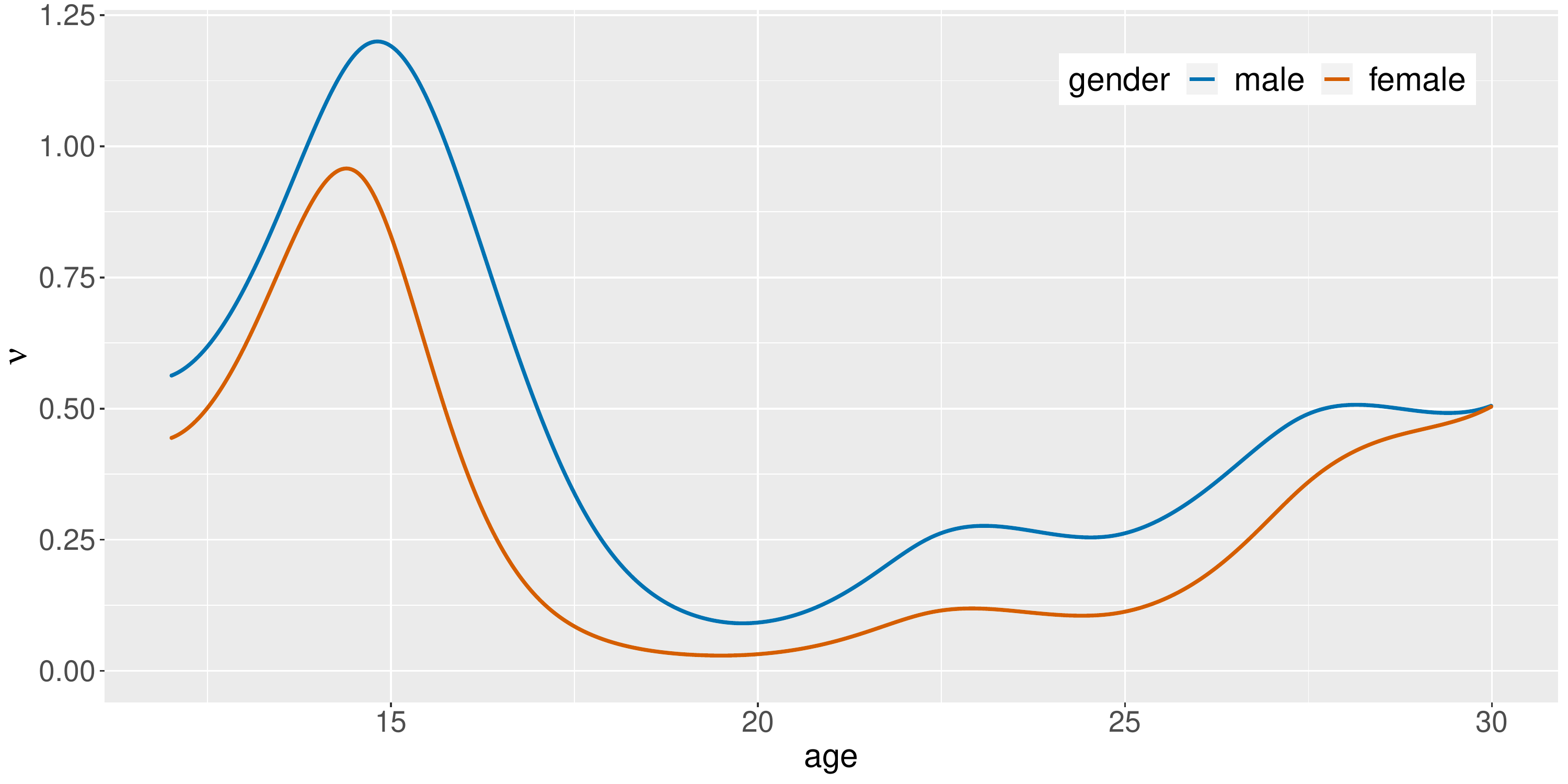}
    \caption{Estimated effect of age on the expected number of offences for male (blue) and female (red) adolescents, respectively, given that the state equals 0.}
    \label{fig:splineEffect}
\end{figure}

The estimated effects of age and gender on the mean parameter of the negative binomial distribution are visualised in Figure \ref{fig:splineEffect}.
While the effect of age on the expected number of offences is quite similar for both gender, female adolescents generally display a lower level of delinquent behaviour than males, which corresponds to the current state of research \citep[e.g.][]{reinecke2013}.
Overall, the effect of age is highly nonlinear. 
Until the age of 14 to 15, there is an increase in delinquent behaviour, followed by a steady decline in the expected number of offences, which reflects the typical age-crime curve \citep[e.g.][]{moffitt1993}.
During the twenties, the expected number of offences increases again, which might here mainly be caused by data collection issues as young adults can commit additional offences that are not considered for adolescents. 

\begin{figure}[h]
    \centering
    \includegraphics[scale=0.4]{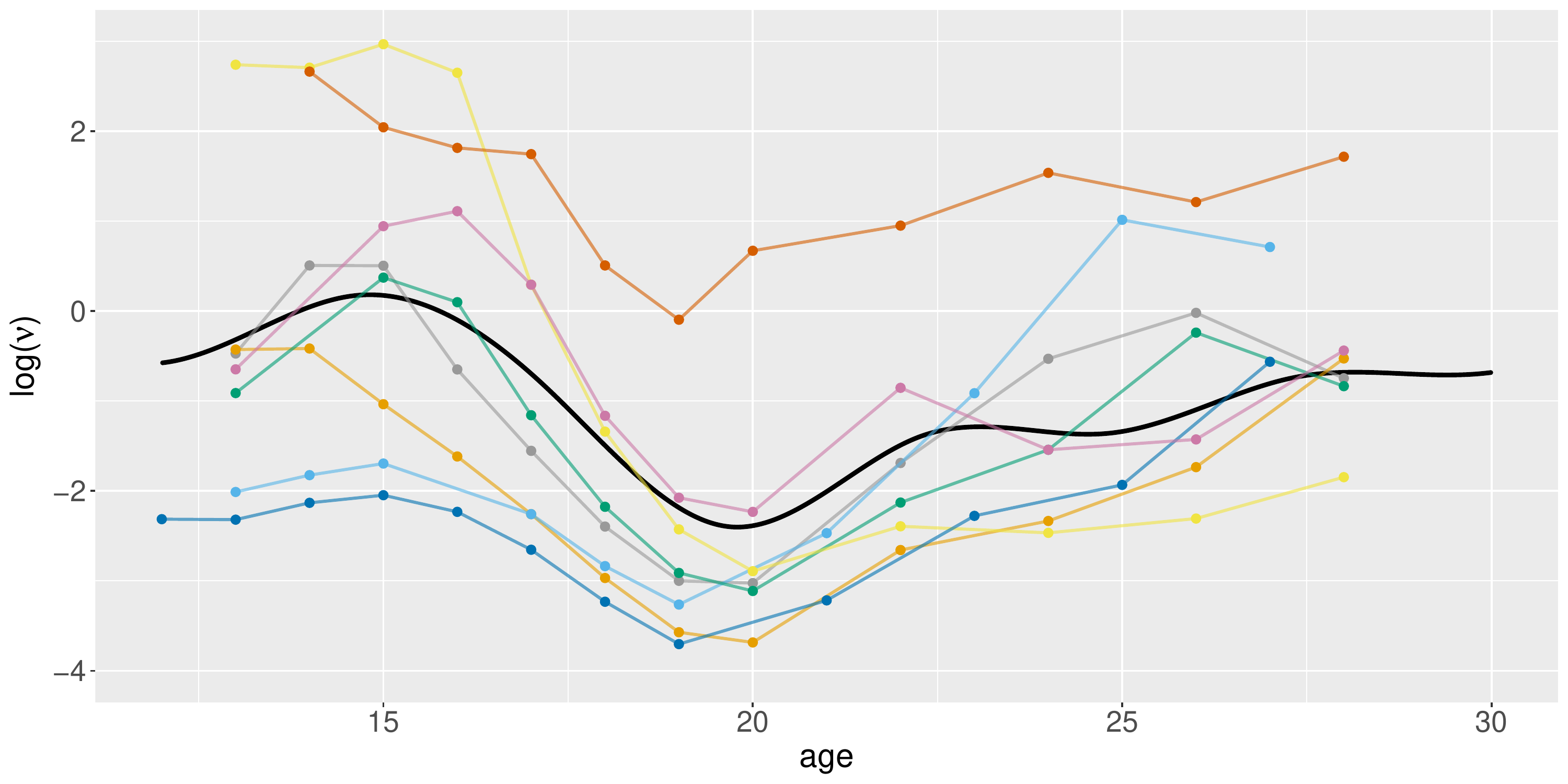}
    \caption{Example trajectories of the logarithm of the expected number of offences for eight male individuals based on their decoded delinquency level at each observation time. The thicker, black line represents the expected trajectory for male adolescents, when their delinquency level is in equilibrium.}
    \label{fig:decStates}
\end{figure}

Due to transferring the SSM to an HMM framework (cf.\ Section \ref{sec:methods}), we can gain additional insight into the delinquency levels of individuals by using the Viterbi algorithm to infer the most probable sequence of underlying states.
Based on these decoded delinquency levels as well as the individuals' gender and age, the expected number of offences can be calculated at each observation time.
Such decoded trajectories are shown for eight male adolescents in Figure~\ref{fig:decStates}. 
As a result of the underlying delinquency levels, individuals' trajectories of the expected number of offences deviate from the overall age trend and fluctuate around the latter.
Moreover, different trajectories are visible: while some adolescents have a permanently increased or reduced level of delinquency, others show early or late periods of increased delinquency levels.

\section{Discussion}
\label{sec:discus}

In this contribution, we developed a flexible framework for formulating and estimating general continuous-time SSMs. 
These are latent-state models suited to sequential observations that are irregularly spaced in time, i.e.\ data to which discrete-time models are not (directly) applicable. 
In some applications, for example in biology \citep[e.g.][]{runde2020}, psychology \citep[e.g.][]{deHaan2017}, or finance \citep[e.g.][]{kim2008}, irregularly spaced observations are simply treated as if they do follow a regular sampling scheme, or are forced into a sequence with regular (i.e.\ equidistant) time intervals based on data aggregation or imputation. 
These aggregated or imputed data are then analysed using discrete-time models, which are less technically challenging than their continuous-time counterparts. 
However, temporal aggregation of continuous-time processes discards information on the exact observation times and introduces subjectivity concerning the choice of the discrete-time modelling resolution, while imputation methods for generating regular time intervals introduce additional uncertainty, which is why both approaches possibly produce biased estimates \citep[see, e.g.,][]{yip2002, delsing2005, barbour2013, kleinke2020}.
Therefore, continuous-time models are generally preferable when data are collected at irregular points in time. 
These models are not only conceptually appealing as their interpretation does not depend on the time resolution of the data at hand, but also avoid the pitfalls mentioned above. 
These benefits come at the cost of increased mathematical and computational complexity, especially for the case of SSMs with non-linear and non-Gaussian processes. 

While we are not the first to consider continuous-time SSMs, existing models often focus on a particular data application and hence are very case-specific \citep[e.g.][]{dennis2014, albertsen2015, niu2016}. 
In particular, existing approaches usually make restrictive model assumptions to simplify parameter estimation, for example requiring the SSM to be linear and Gaussian to enable the application of the Kalman filter \citep[e.g.][]{johnson2008, tandeo2011, koopman2018, lavielle2018, jonsen2020}. 
In contrast, the maximum (approximate) likelihood approach we propose here is not tied to specific distributional or linearity assumptions, thus allowing for both non-linear and non-Gaussian specifications of the state and observation process. 
Our method, however, is by no means the only method to fit continuous-time SSMs: 
Apart from the Kalman filter, which can be used for linear and Gaussian SSMs, MCMC methods \citep{niu2016} and Laplace approximation techniques \citep[][]{albertsen2015, michelot2020} as implemented in the R-package Template Model Builder \citep[][]{Kristensen2016} have been developed for statistical inference in continuous-time SSMs.
While the modelling approach presented here is not assumed to be superior to such alternative estimation techniques, it offers the convenience of the continuous-time HMM framework and its corresponding efficient algorithms. 
The latter proves beneficial not only with respect to model fitting but also for decoding the most probable underlying state trajectories. 
Moreover, only minor changes in the corresponding code for the likelihood calculation are required to consider different distributions or non-linear relationships in either the observation or state process, provided that the transition density is known in explicit form. 
A major caveat of the approach, however, is that it suffers from a curse of dimensionality when considering multivariate state processes \citep[e.g.][]{langrock2011}. 
In conclusion, our approach constitutes an accessible and very flexible framework for modelling irregularly spaced sequential data driven by a one-dimensional underlying state process.

\section*{Acknowledgements}

We would like to thank Christiane Fuchs for her valuable input on SDEs.

\newpage

\appendix

\counterwithin{figure}{section}

\section{Appendix}

\subsection*{Additional information on the simulation experiments}

\begin{figure}[!htb]
    \centering
    \includegraphics[width=150mm]{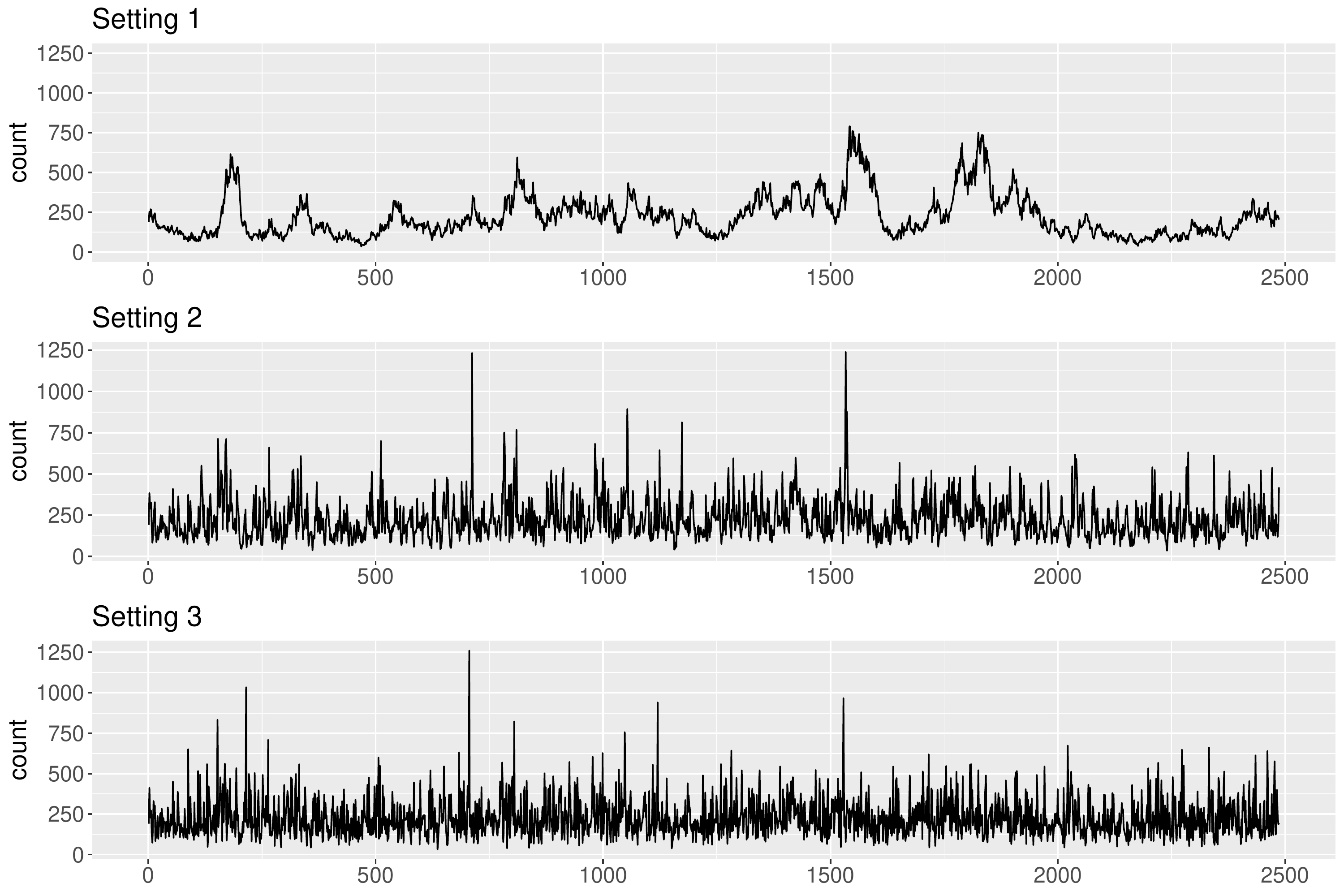}
    \caption{Simulated time series of irregularly spaced observations for each of the three simulation settings described in Section~\ref{sec:sim}. The observations are generated by the same observation process, while the parameters of the underlying state process differ between the settings.}
    \label{fig:simData}
\end{figure}

\begin{figure}[!htb]
    \centering
    \includegraphics[width=150mm]{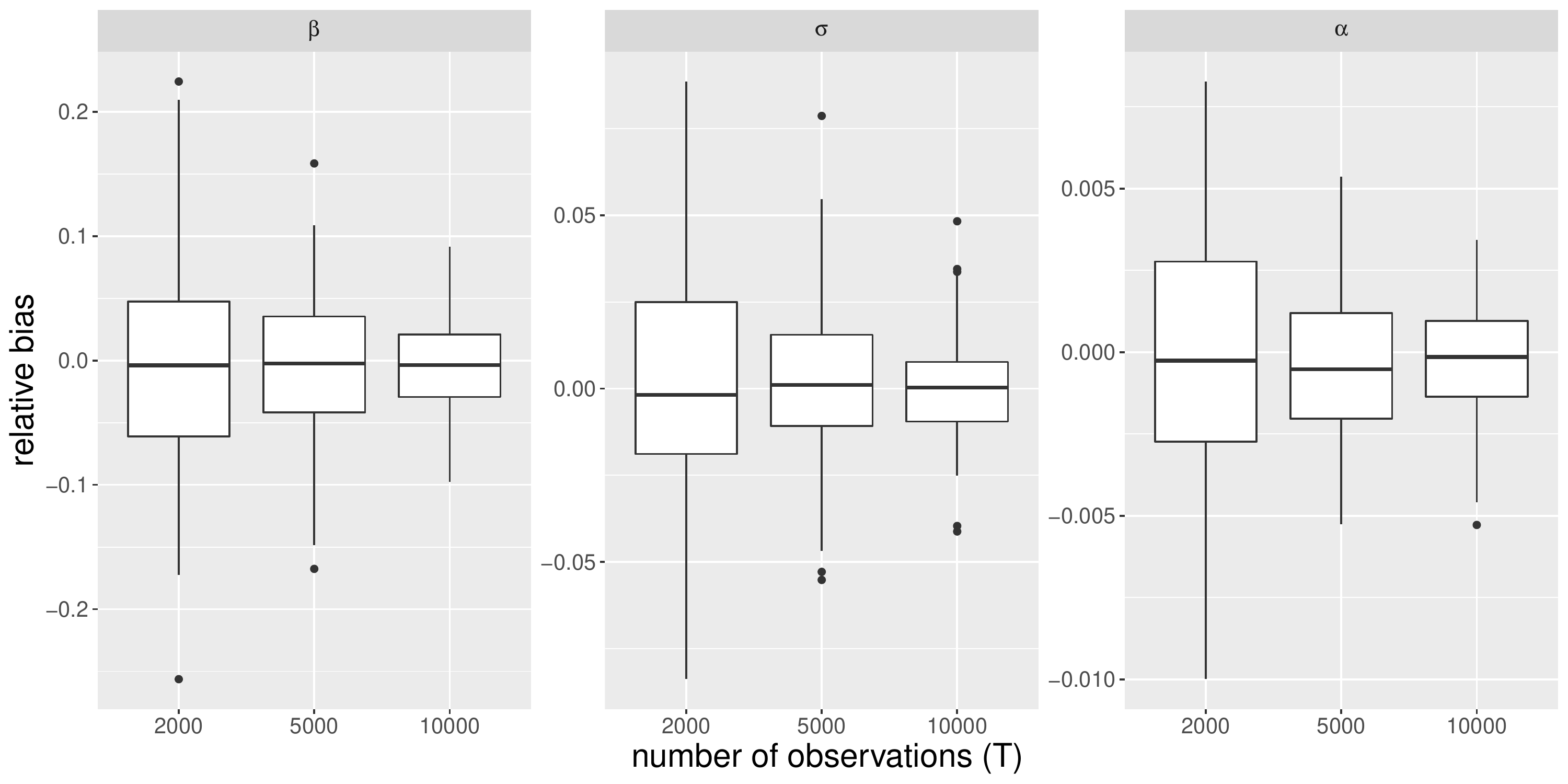}
    \caption{Boxplots of relative bias of the estimated model parameters from 200 simulation runs for $T = 2000, 5000, 10000$ observations. True parameter values are $\beta = 0.5$, $\sigma = 0.5$, and $\alpha = 200$.}
    \label{fig:bias}
\end{figure}

\newpage

\subsection*{Additional information on the results of delinquent behaviour in adolescence and young adulthood}

\begin{figure}[!htb]
    \centering
    \includegraphics[width=140mm]{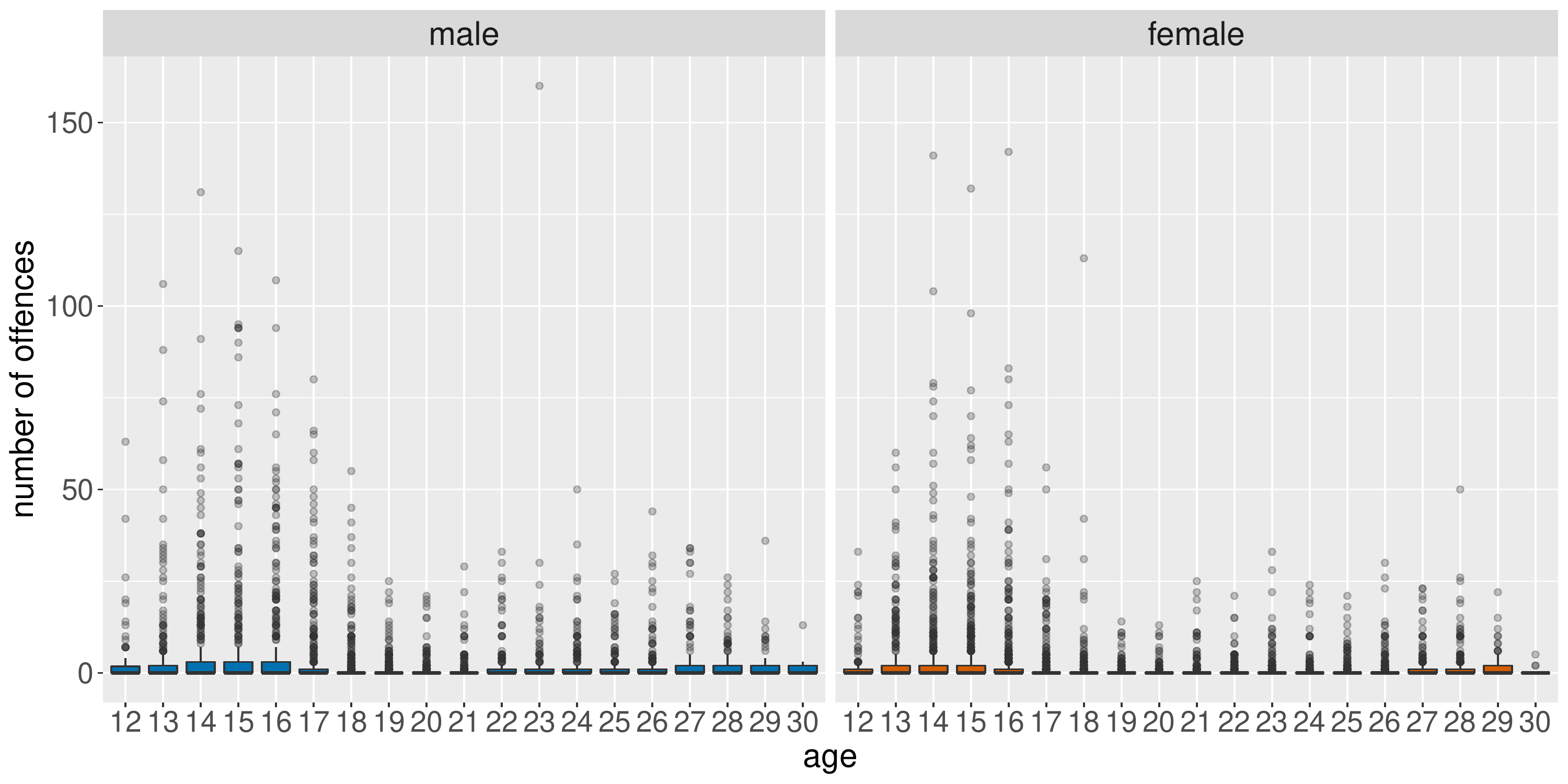}
    \caption{Boxplots of the number of offences committed in the twelve months prior to the survey for different age classes and both gender.}
    \label{fig:boxplots_outliers}
\end{figure}

\begin{figure}[t!]
    \centering
    \includegraphics[width=130mm]{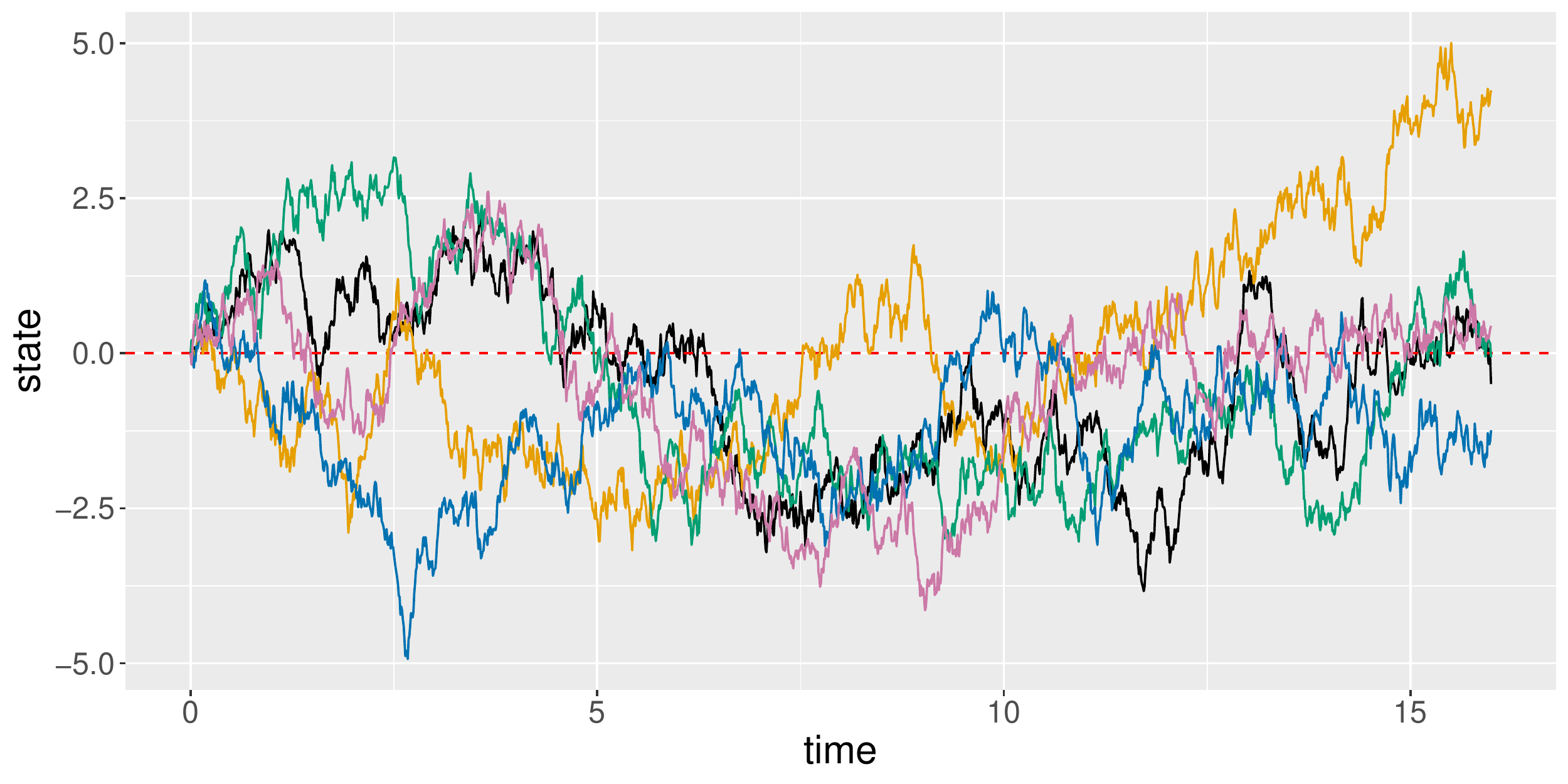}
    \caption{Simulation of possible state trajectories for the study period of 16 years based on the estimated parameters of the OU process. The red dashed line indicates the intercept around which the processes fluctuate. The graphs have been obtained by application of the Euler-Maruyama scheme with initial value 0 and step length 0.01.}
    \label{fig:OUresults}
\end{figure}

\end{spacing}

\end{document}